\documentclass[11pt]{article}

\input epsf.sty

\def\lromn#1{\uppercase\expandafter{\romannumeral#1}}

\begin{document}

\begin{flushright}
\end{flushright}

\begin{center}
\begin{Large}
\textbf{
A new method of creating high intensity neutron source
}
\end{Large}

\vspace{2cm}
T. Masuda, A. Yoshimi, and
M.~Yoshimura

\vspace{0.2cm}
Research Institute for Interdisciplinary Science,
 Okayama University \\
Tsushima-naka 3-1-1 Kita-ku Okayama
700-8530 Japan

\end{center}

\vspace{3cm}

\begin{center}
\begin{Large}
{\bf ABSTRACT}
\end{Large}
\end{center}

We propose a new scheme of producing intense neutron
beam whose yields exceed  those of existing facilities by
many orders of magnitude.
This scheme uses the recently proposed
photon beam extracted from circulating quantum ions,
which is directed to a deuteron target for photo-disintegration.
The calculated neutron energy spectrum is nearly
flat down to neV range, except a threshold rise and
its adjacent wide structure.
Hence, there exists a possibility of directly using  sub-eV
neutrons   without a moderator.
We shall have brief comments on promising 
particle physics applications using this large yield of neutron.

\vspace{4cm}

Keywords
\hspace{0.5cm} 
Neutron source,
Heavy ion synchrotron, 
Quantum coherence,
Gamma ray beam,
Photo-disintegration of  deuteron,
Time-reversal invariance,
Ultra cold neutrons

\newpage


\section
{\bf Introduction}

Neutron is a powerful source to explore  fundamental  physics
and is an indispensable tool in applications to material and life sciences.
Despite of existing strong neutron sources using
reactor and spallation  facilities  \cite{existing facilities} new ideas 
of more intense sources \cite{proposals of neutron source}
are obviously welcome from this point.

We propose in the present work a new scheme of accelerator-based neutron
source.
In addition to a potentiality of producing neutron yield
much stronger than existing sources,
this scheme creates a neutron spectrum
calculable from the first principles and
it is possible to use this source without
using  moderator \cite{moderator}.
The radiation safety and the heat generation problems appear less severe compared with
existing methods.

For definiteness we consider an example to use the parameter of an
existing accelerator,  the SuperKEKB ring \cite{superkek}.
Hence we take a circumference of the ring $\sim 3\,$km,
and an ion boost factor in the range of $100 \sim 300$.

\begin{figure*}[htbp]
 \begin{center}
 \epsfxsize=1.0\textwidth
 \centerline{\epsfbox{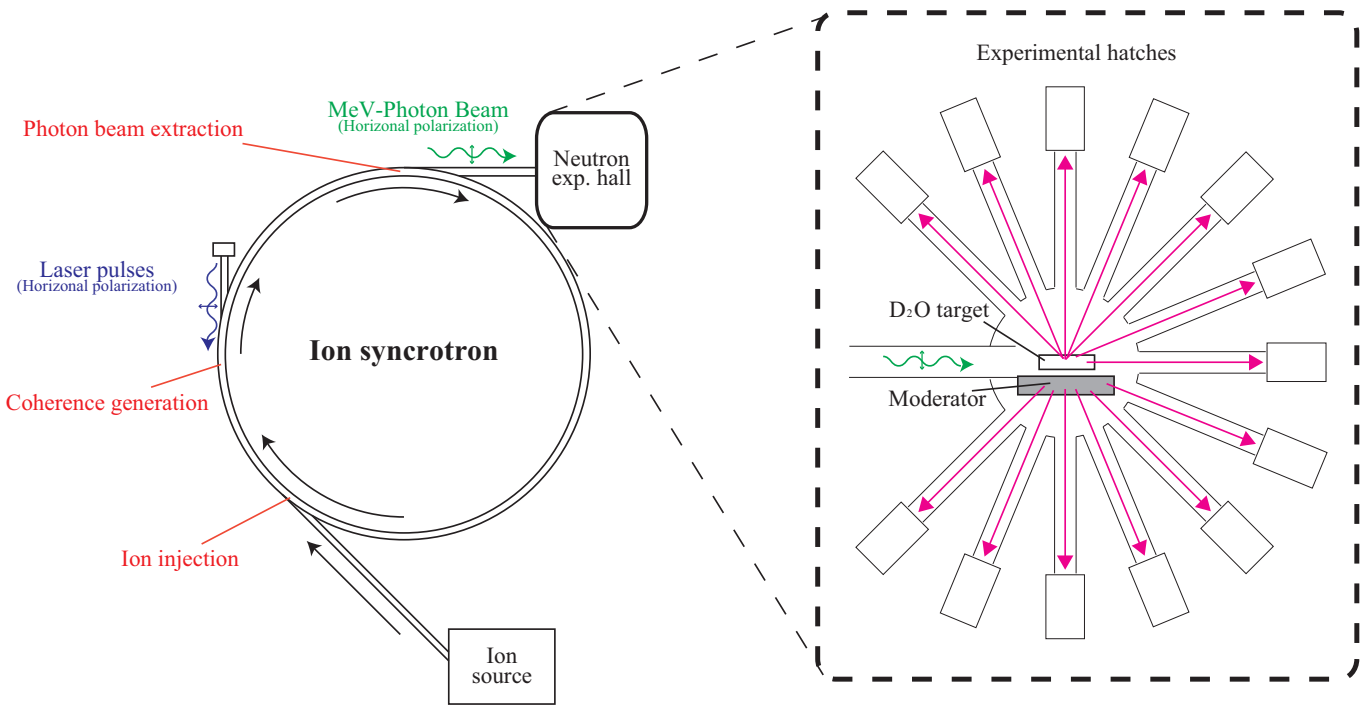}} \hspace*{\fill}
   \caption{
Schematic view of high intensity neutron beam facilities.
}
   \label {schematic of neutron beam facility}
 \end{center} 
\end{figure*}

Our proposed idea of neutron source is depicted schematically in 
Fig(\ref{schematic of neutron beam facility}).
It consists of an ion synchrotron that produces an intense and
well-collimated gamma ray beam based on \cite{pair beam},  \cite{new pair beam}, 
a target area that produces neutrons
by photo-disintegration of deuteron and a moderator system that
transports neutrons to experimental sites.
The option of direct slow neutron extraction without the use of moderator
is possible.

In the rest of this work
we first discuss essential features of the photon beam
from circulating ions at the synchrotron, and
proceed to calculation of the neutron yield at the target.
A brief discussion of particle physics applications that may
become promising with high intensity neutron yield is added.

Throughout this work we use the natural unit of $\hbar = c = 1$.

\vspace{0.5cm} 
\section
{\bf Photon beam from quantum  ion in circulation}

It was recently proposed \cite{pair beam},  \cite{new pair beam}
that quantum  ions in circulation
emit strong photon beam.
The beam intensity depends on the coherence $\rho_{eg}$
of quantum ions,
which is defined for a state of quantum mixture,
\begin{eqnarray}
&&
| c(t) \rangle = \cos \theta_c  |g \rangle 
+ \sin \theta_c e^{-i \epsilon_{eg} t/\gamma}  | e\rangle
\,, \hspace{0.5cm}
\rho_{eg} =   \cos \theta_c \sin \theta_c
\,.
\end{eqnarray}
The coherence given by $\rho_{eg}$ is generated by irradiation of lasers
from counter-propagating directions.
In ordinary synchrotron radiation $\rho_{eg} = 0$ and
the gamma ray emission discussed here is different from
the synchrotron radiation.
We use linearly polarized lasers in the ion orbit plane
to select the dominant direction of neutron flux within the plane.
Ion candidate state $|e\rangle$ (the ground state for $|g \rangle$) 
we consider here
are  He-like 2$\,^3$S state \cite{he-like a}, \cite{text book on atoms},
but there may be many other possibilities.

\begin{figure*}[htbp]
 \begin{center}
 \epsfxsize=0.7\textwidth
 \centerline{\epsfbox{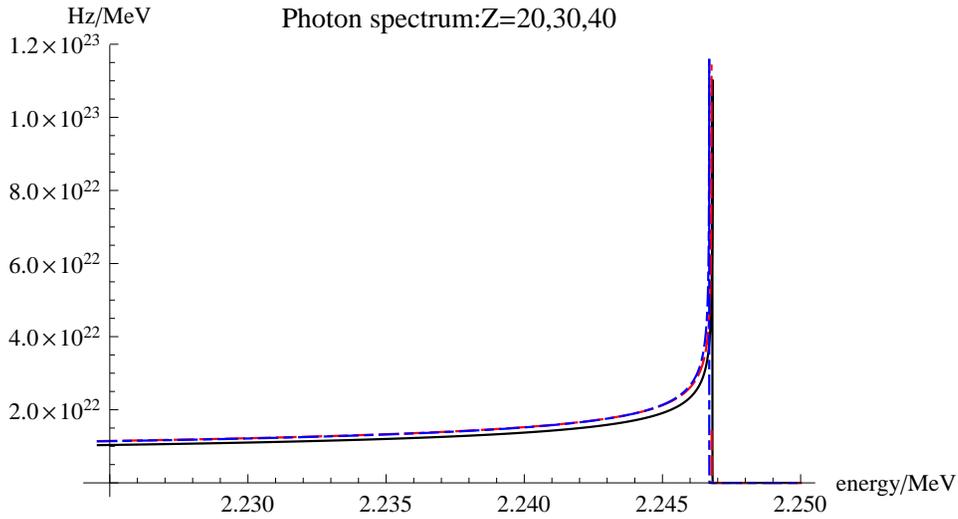}} \hspace*{\fill}
   \caption{ 
Photon  spectrum above the photo-disintegration threshold,
photons being emitted in the forward direction after extraction at
ion turn of angle $\pi/2$ after laser irradiation. 
He-like ions of atomic number $Z$ and the boost factor $\gamma$  taken are 
$Z(\gamma) =$ 20 (274)
 in solid black, 30 (121) in dashed red, and
40 (67.5) in dash-dotted blue.
The boost factor $\gamma$ here is taken to 
satisfy the condition $\omega_m \sim 2\gamma \epsilon_{eg} $ equal to
the photo-disintegration  threshold $S_n \times (1+ 0.01) $
(1 \% away from the disintegration threshold).
Relaxation width 1 eV, the number of ions $N_I = 10^{9}$,
the initial coherence $\rho_{eg}(0) = 1/10^3$
(corresponding to the effective available ion number $N_I \rho_{eg}^2(0) = 10^3$)
are assumed.
}
   \label {g energy spectrum 1}
 \end{center} 
\end{figure*}

The emitted photons are limited 
by the solid angle area $\pi/\gamma^2$ near
the forward direction to the circulating ion.
The forward rate  is given by
\begin{eqnarray}
&&
(\frac{ d^2 \Gamma_{\gamma} }{ d x d\Omega})_0 = 
\frac{  A_{eg}}{ 4 \sqrt{\pi} } N_I \rho_{eg}^2(t) \sqrt{\rho \epsilon_{eg}} 
\frac{\gamma}{\sqrt{\beta}}
\frac{1}{x}
 \left( \beta^2  x^2 - ( x - \frac{1}{\gamma} )^2
\right)^{-1/4} 
\,,
\label {forward g rate}
\end{eqnarray}
using the dimensionless energy $x = \omega/\epsilon_{eg}$.
The quantities introduced here are $\rho$ the radius of the ring,
$N_I $ the  number of ions in the ring,
$\gamma = 1/\sqrt{1-\beta^2}$ the boost factor,
and $A_{eg}$ the A-coefficient (decay rate)  for 
$| e\rangle \rightarrow |g\rangle + \gamma$ of the level spacing $\epsilon_{eg}$.
Photon beam intensities, as illustrated in Fig(\ref{g energy spectrum 1})
taking into account the  de-coherence discussed below,
are much stronger than any presently available photon beams.

It is important to estimate the coherence loss \cite{coherence loss}
in order to determine where in the ring the extraction of the photon beam is made.
The basic equation of the time dependence  and its solutions for the coherence loss is
\begin{eqnarray}
&&
\frac{d \rho_{eg}} {dt} = - \frac{G}{2} \rho_{eg}^3 
\,, \hspace{0.5cm}
\rho_{eg}(t) = \frac{\rho_{eg}(0)} {\sqrt{1+G\rho_{eg}^2(0) t }}
\,, \hspace{0.5cm}
\rho_{eg}(0) = \frac{1}{2} \sin (2\theta_c)
\,,
\\ &&
G = \frac{A_{eg}}{4\sqrt{\pi}}  \sqrt{\rho \epsilon_{eg}}
\frac{\pi}{\gamma \sqrt{\beta}}\int dx 
\frac{1}{x} (\beta^2 x^2 - (x - \frac{1}{\gamma})^2)^{-1/4}
\,,
\end{eqnarray}
where $G$ was calculated by integrating the photon number over
all  emitted photon energies and angular area $\pi/\gamma^2$.
It is interesting to note that
the asymptotic value in $t \gg 1/(G \rho_{eg}^2(0)\,)$ is independent of the initial coherence $\rho_{eg}(0)$,
$\rho_{eg}(t) \rightarrow 1/\sqrt{Gt}$.

We choose in the rest of calculation  a parameter  $\rho_{eg}(0) $ to be $ 1/10^3$, 
which is controllable by laser irradiation, and
$f$ to be around $1/4$ corresponding to the extraction
point at 3/4 km of the SuperKEKB ring.
We assume throughout the present work
that the circulation velocity $\beta$ is constant despite of an ion energy loss
caused by photon emission.
It is implicitly assumed here that the ion energy is compensated by its
acceleration, thereby justifying the assumption of the constant velocity
approximation.
We take in the present work He-like ions using ``forbidden'' $2\,^3 S \rightarrow 1S$.
Transitions of this type  actually occur for  high atomic number $Z$,
their decay rates scaling with $Z^{10}$ \cite{text book on atoms}.
Using theoretical data of \cite{he-like a}, 
we numerically fit the A-coefficient of He-like transitions, 
and use the level spacing derived from the relativistic Dirac equation.
In presented figures of this work we took as an illustration
He-like ions of Ca$^{18+}$, Zn$^{28+}$, Zr$^{38+}$.
We include a relaxation width effect by
multiplying the photon energy spectrum at $\omega_0$
by a convolution factor $\Delta /(\,(\omega_0 - \omega)^2 + \Delta^2/4)$ 
to be integrated over $\omega_0$.
The relaxation width $\Delta$ is related to photon emission  via $G \rho_{eg}^2(t)$,
but other effects can give much larger width factor.
Since results are insensitive to the actual value,
we assume $\Delta = 1$eV in our numerical computations.
A resultant  photon spectrum  is illustrated in Fig(\ref{g energy spectrum 1}).

\vspace{0.5cm} 
\section
{\bf Calculation of neutron yield}

Emitted photon beam is directed to the target area
where photo-disintegration occurs to produce neutrons.
We place the target not too far from the extraction point of
photon beam in order to be fully covered by the angle area within $\pi/\gamma^2$.
Deuteron photo-disintegration,
$\gamma + D \rightarrow n + p$, have been
analyzed both experimentally   \cite{utsunomiya}  and theoretically
\cite{segre}, \cite{marshall-guth} in great detail.
They show that electric dipole (E1) contribution
giving the angular distribution $\sin^2 \theta_n (1+ \cos (2\varphi_n)\,)$
($\theta_n$ is the neutron emission angle from the beam direction, while $\varphi_n$
the angle out of the orbit plane)
is dominant away from the threshold $\sim 2.2$\,MeV, while the isotropic
magnetic dipole (M1) contribution is important near the threshold.
The differential photo-disintegration cross section we use
(the photon energy $\omega$ given in MeV unit) is 
\begin{eqnarray}
&&
\frac{d^2\sigma_{\gamma {\rm D}}}{d\omega d\Omega} = 
\left( \frac{d\sigma_{\gamma  {\rm D}}}{ d\Omega}\right)_0
\delta (\omega + m_{ {\rm D}} - K_n -\sqrt{ m_p^2+ \omega^2+p_n^2 - 2\omega p_n \cos\theta_n })
\,,
\\ &&
\left( \frac{d\sigma_{\gamma {\rm D}}}{ d\Omega}\right)_0 =
\left(
62. 77 \frac{(\omega - S_n)^{3/2} }{\omega^3} \frac{3}{8\pi} \sin^2 \theta_n
+ 0.692 \frac{\sqrt{ \omega - S_n}} {\omega (\omega - 2.15)} \frac{1}{4\pi}
\right)
\times 10^{-27}\, {\rm cm}^2
\,,
\end{eqnarray}
taking a numerical fit that includes both E1 and M1 contributions.
The deuteron binding energy is denoted by $S_n$ which is  2.22457
MeV \cite{deuteron be}. 
The neutron kinetic energy $K_n = p_n^2/2m_n$ is related to the emission angle $\theta_n$
due to the two-to-two body reaction.

Using the cross section of photo-disintegration 
given as a function of $(\omega, \cos \theta_n)$,
the neutron rate emitted at an angle $\theta_n$ measured in the ion orbit plane from
the photon beam is given by
\begin{eqnarray}
&&
\left(\frac{d^2 \Gamma}{dK_n d\cos \theta_n } \right)_{\varphi_n =0} = n_{ {\rm D}} L
\sqrt{\frac{2 K_n}{m_{\rm N}}}
\left(
\left( \frac{d\sigma_{\gamma {\rm D}}}{ d\Omega}\right)_0
\Delta \Omega (\frac{ d\Gamma_{\gamma} }{dx})_0
\right)_{\omega = X(K_n, \cos \theta_n )}
\,,
\\ &&
X (K_n, \cos \theta_n) = 
\frac{1}{2} \frac{S_n(2m_{\rm N} - S_n) +2m_{\rm D} K_n }{m_{\rm N}- S_n - K_n + \sqrt{2m_{\rm N} K_n} \cos \theta_n }
\,.
\end{eqnarray}
The kinematic relation, $\omega = X(K_n, \cos \theta_n )$, 
among the photon energy $\omega$,
the deuteron emission angle $\theta_n$ and its kinetic energy $ K_n$, 
proper to the two-body process was used.
The quantity $n_{\rm D} L$ is the deuteron number density per target cross area.

An advantage of the proposed scheme is that
one can optimize the photon beam by
a choice of accelerator parameters, $Z$ and $ \gamma$.
Choose the maximum photon energy  $ \omega_m \sim 2 \gamma \epsilon_{eg}$
to be near the threshold like 
$S_n < \omega_m \leq S_n (1 + \epsilon )$.
For He-like ions this reads roughly as
\begin{eqnarray}
&&
\frac{S_n}{20\, {\rm eV}} < \gamma Z^2 \leq \frac{S_n}{20\, {\rm eV}} (1 +\epsilon)
\,.
\label {photon energy tuning}
\end{eqnarray}
Note that the required boost factor $\gamma$ is at least larger than
$S_n/(Z^2 20\, {\rm eV}) \sim 10^3 (10/Z)^2$.
There exists a kinematic restriction:
$\epsilon \geq S_n(1-\cos^2 \theta_n)/(2m_{\rm N})$.

For numerical calculations we assume a liquid D$_2$O target of
small cross sectional area of order 1\,cm$^2$ 
transverse to the photon beam.
In this type of targets one
may ignore neutron scattering within the target D$_2$O.
Furthermore, 
the effect of Compton scattering of the beam photon off atomic electrons 
can be taken into account by a simple beam
reduction factor, $e^{- \mu z},$ with 
$\mu$ the attenuation coefficient \cite{attenuation coef}
(dominated by the Compton scattering, hence $\mu \sim \sigma_C n_e$)
and $z$ the target location from the beam entrance.
We assume for the target $1\, {\rm cm}^2 \times 20\,$cm D$_2$O
to use the beam reduction factor
of $\int_0^{\infty} dz  e^{-\mu z} \sim 20$ cm
(giving the total deuteron number $\sim 1.2 \times 10^{24}$).
The circulating ion number is taken as $N_I =10^{9}$ 
($1/10^4$ of the number taken from that achieved for accelerated protons in TeV region),
and the initial coherence as
$\rho_{eg}(0) = 1/10^3$.
With increased $N_I$ the event rate linearly increases.
The neutron yield given in the figures refers to
the value per unit solid angle area at the given neutron emission angle.

\begin{figure*}[htbp]
 \begin{center}
 \epsfxsize=0.6\textwidth
 \centerline{\epsfbox{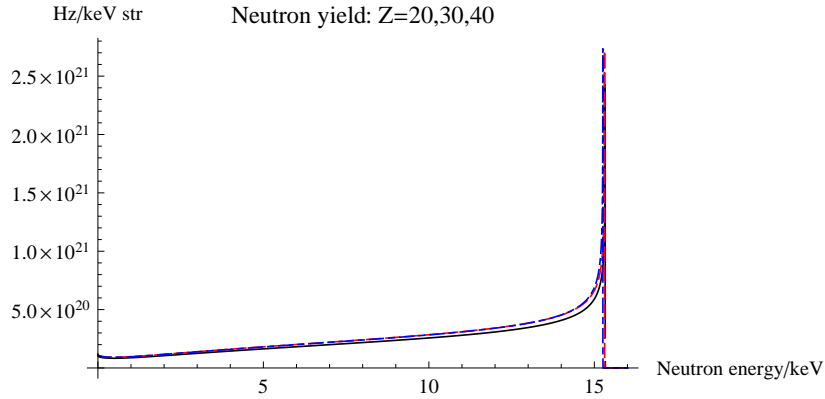}} \hspace*{\fill}
   \caption{ 
Neutron yield  at an emission angle $\pi/6$ from the photon beam.
The same set of $(Z, \gamma)$ combination as in Fig(\ref{g energy spectrum 1})
is taken.
Relaxation width 1 eV, the number of ions $N_I = 10^{9}$,
the initial coherence $\rho_{eg}(0) = 1/10^3$.
the deuteron number $n_{ \rm D} L = 1.3 \times 10^{24}\,$cm$^{-2}$
(roughly corresponding to 20 g per 20\,cm target length) are assumed.
}
   \label {neutron flux overall}
 \end{center} 
\end{figure*}

\begin{figure*}[htbp]
 \begin{center}
 \epsfxsize=0.6\textwidth
 \centerline{\epsfbox{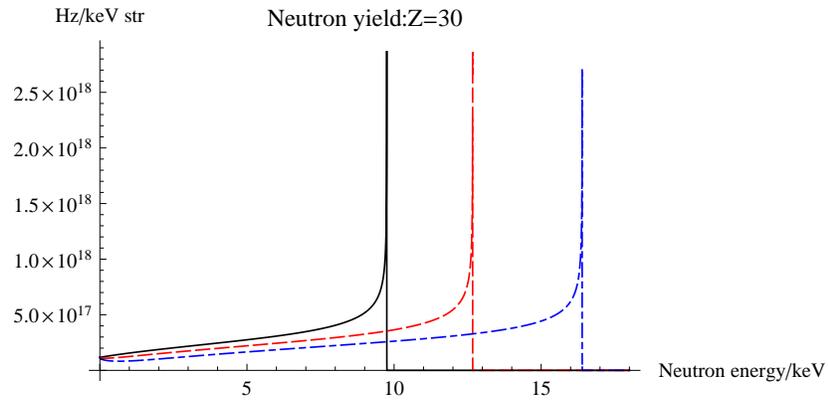}} \hspace*{\fill}
   \caption{ 
Neutron yields extracted at various angles:
$\pi/2$ in solid black, $\pi/3$ in dashed red, and
0 (forward direction) in dash-dotted blue.
Other parameters are taken the same as in Fig(\ref{neutron flux overall}).
}
   \label {neutron flux at angles}
 \end{center} 
\end{figure*}

\begin{figure*}[htbp]
 \begin{center}
 \epsfxsize=0.6\textwidth
 \centerline{\epsfbox{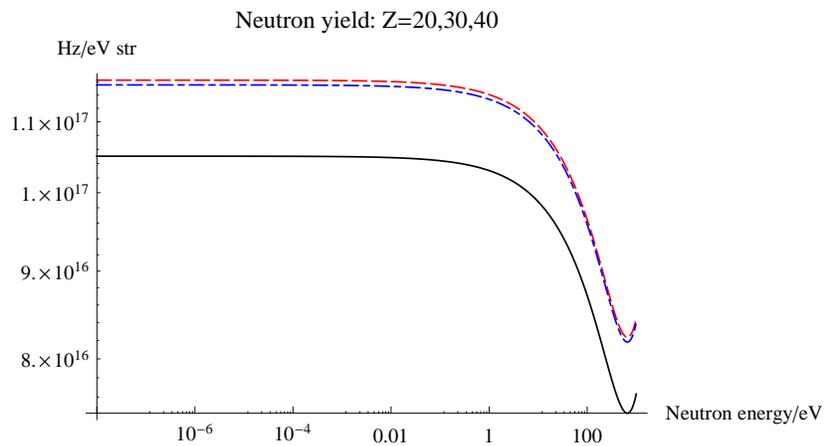}} \hspace*{\fill}
   \caption{ 
Neutron yield including  the sub-eV range  at angle $1/10^3$.
Parameters are the same as in Fig(\ref{neutron flux overall}).
}
   \label {neutron flux subev}
 \end{center} 
\end{figure*}

The global feature of the neutron yield is shown
in Fig(\ref{neutron flux overall}), which shows 
yields much larger than those of exiting facilities:
the yield numbers integrated over the entire neutron
energy are  to be compared with the designed J-PARC MLF  value,
which is of order, $10^{17}/4\pi\,$ Hz,  at the target point.
Neutron yields extracted at different angles have different maximum
energy cutoffs, as shown in Fig(\ref{neutron flux at angles}).
The total 100 neV range of neutron yield using the parameter set in the figures 
is $O(10^{10})\,$Hz/100 neV for $Z=O(30)$, as evident in Fig(\ref{neutron flux subev}).
This number is $O(10^{14})$Hz/meV in the meV range.
These large yields should be a great benefit to
fundamental physics and other applications.
The conventional use of moderators reduces the flux at
target by an order of $10^3$, while this scheme gives a nearly flat spectrum
down to neV range.

The large and flat yield  shown here opens a new possibility of using directly the sub-eV neutron
 without a moderator.
The example of Fig(\ref {neutron flux subev})
uses the boost factor fine tuned to
$S_n (1 + 0.01)/20\, {\rm eV}$.
A  great advantage of our new scheme is
that one can adjust the parameter $\epsilon$ of eq.(\ref{photon energy tuning})
to precisely select the neutron energy region both from above and
below, since the maximum photon energy $\sim 2 \gamma \epsilon_{eg}$ is
fixed by acceleration of ions and
the minimum usable photon energy $\sim $\,2.2\,MeV is determined
from the threshold of deuteron photo-disintegration.

\vspace{0.5cm} 
\section
{\bf Summary and prospects}

We proposed a method of how the intense photon beam produced from
circulating quantum ions  can provide high intensity neutron
yields much stronger than those of existing facilities.
The neutron flux thus obtained is pulsed in correlation with the bunch
structure of circulating ions and timing of laser irradiation.
Its calculated spectrum is nearly flat down to the neV range.

The success of high intensity neutron source project
rests with R and D works on the photon beam
in which it is important to realize a large value
of parameter combination $N_I \rho_{eg}^2(0)$.
We have assumed in our sample calculations that the total available ion number  $N_I $
is $ 10^{9}$, taking a value, $1/10^4$ 
of the achieved value from accelerated protons in the TeV region,
and the initial coherence $\rho_{eg}(0) = 1/10^3$ which
is controllable by laser irradiation.
These values may or may not be  optimistic for assumed ions, since one has to accelerate
He-like ions in the intermediate $Z$ range.
If this choice of parameter set is too optimistic,
one may attempt to move the position of extraction point
closer to the laser irradiation point in order to
obtain a higher gamma ray intensity.
There may be other choices different from He-like ions.
We have also studied the simple system of H-like E1 allowed
$2\,^1P_1 \rightarrow 1\,^1S_0$
transitions. This system gives $\sim 0.3$ reduction of rates  around $Z=30$ in comparison with He-like
system studied above.
For more complicated system of ions 
 a systematic atomic physics calculation
is required to derive basic data of A-coefficients and level spacings.
The systematic study of the optimal choice of the ion atomic number and
the boost factor  is clearly important for further development.
As a target of photo-disintegration we considered deuteron
in the present work.
Another possibility worth of investigation is  solid $^9$Be target.

The rest of downstream facilities such as the
target and the moderator is straightforward.
Conventional moderator systems are useful for experiments employing thermal neutrons. 
If one wants to directly use  sub-eV neutrons which are abundant
at the deuteron target in the neutron source of our scheme,
one  may need construction of
 a reliable system of energy separation or a velocity selector.

With realization of high intensity neutron source,
both fundamental physics and applications to
material and life sciences may have brighter future.
In particle physics one may list interesting applications with some comments.
1. UCN (Ultra Cold Neutrons). 
One stores neutrons below $\sim 300$ neV in totally reflecting bottles,
for instance in search for electric dipole
moment (EDM) of neutron whose presence indicates violation
of the fundamental symmetry, time-reversal symmetry.
Our integrated rates are of order $10^{10}$ Hz/(str 100 neV).
This may give a large storage integrated over
a fraction of neutron lifetime.
2. T-odd triple product among the beta decay products of neutrons.
A precise selection of
the parent neutron momentum $\vec{p}_n$ should help to search for
a few types of T-odd observables, its presence  indicating violation of time-reversal invariance.
Precision experiments should be possible with
a large neutron flux $O(10^{14})\,$Hz/(str meV).
Discovery of a finite triple product of this combination may indicate 
new physics beyond the standard electroweak theory,
since there is a wide gap between the present upper limit
and expectation of the standard theory.
We refer to references \cite{neutron science} on other applications.
In a long run one is tempted to use
this intense neutron source for resolution
of the nuclear waste problem by nuclear transmutation.

\vspace{0.5cm}
 {\bf Acknowledgements}

 One of us (M.Y.) should like to thank K. Yamada for
a stimulating conversation that led to this investigation.
All of us appreciate for enlightening discussions
H. Shimizu, M. Kitaguchi, K. Hirota and N. Sasao.This research was partially
 supported by Grant-in-Aid for Scientific Research on Innovative Areas
 "Extreme quantum world opened up by atoms" (21104002) from the
 Ministry of Education, Culture, Sports, Science, and Technology.


\begin{thebibliography}{99}
\bibitem{existing facilities}
F. Maekawa et al, Nucl. Instrum. Meth.
 {\bf A620}, 159 (2010).

S. Henderson et al, Nucl. Instrum. Meth.
 {\bf A763},
 610 (2014). 


\bibitem{proposals of neutron source}
A partial list of new proposals for strong neutron sources are

I. Pomernatz et. al, 
Phys. Rev. Lett. {\bf 113}, 184801 (2014).

A. Taylor et al.,  Science{\bf 315} 1092-1095  (2007).

D. Habs et al,
App. Physics {\bf B}
and arXiv:1008.5324v1(2010).



\bibitem{moderator}
M.H. Parajon, E. Abad, F.J. Bernejo, Physics Procedia 60 74-82 (2014).


\bibitem{superkek}
Y. Ohnishi et al,,
Progr. Theor.Exp. Phys.
{\bf 03A} 03A011 (2013).

\bibitem{pair beam} 
M. Yoshimura and N. Sasao,
Phys. Rev. {\bf D92}, 073015(2015)
and arXiv: 1505.07572v2(2015).



\bibitem{new pair beam}
M. Yoshimura and N. Sasao,
{\it Photon and neutrino-pair emission from
circulating quantum ion beam},
arXiv: 1512.06959(2015).


\bibitem{he-like a}
C.D. Lin, W.R. Johnson, and A. Dargarno,
Phys. Rev. {\bf A15}, 154(1977).



\bibitem{text book on atoms}
I.I. Sobelman,
 {\it Atomic Spectra and Radiative Transitions},
2nd edition,
Springer (1992).





\bibitem{coherence loss}
M. Yoshimura, T. Masuda, N. Sasao, and A. Yoshimi,
paper in preparation.

\bibitem{utsunomiya} 
K.Y. Hara et al.,
Phys. Rev. {D 68}, 072001(2003) and references therein.

\bibitem{segre}
E. Segre,
{\it Nuclei and Particles}, 2nd edition, Benjamin (1977).

\bibitem{marshall-guth}
J.F. Marshall and E. Guth,
Phys. Rev. {\bf 78}, 738 (1950).

\bibitem{deuteron be}
M. Wang et al., Chin. Phys. C 36 (2012) 1603.

G. Audi et al., Nucl. Phys. A 624 (1997) 1.



\bibitem{attenuation coef}
M.J. Berger et al,  XCOM.
Photon Cross Sections Database, \\
http://www.nist.gov/pml/data/xcom/index.cfm (2016).



\bibitem{neutron science}
For a review of applications of neutrons to
fundamental physics, see 

D. Dubbers and M.G. Schmidt, 
Rev. Mod. Phys. 83 1111 (2011).




\end{thebibliography}
\end{document}